\documentstyle[psfig]{laa}
\begin{document}
\thesaurus{08.01.1 08.16.3 08.06.3 10.08.1 12.03.3}

\title{ Lithium Abundance of Halo Dwarfs Revised}
\author{  Paolo Molaro$^1$, Francesca Primas$^2$, Piercarlo Bonifacio$^2$}
\offprints{P. Molaro, Osservatorio Astronomico di Trieste, Via G.B.
Tiepolo 11, Trieste, Italy}
\institute{
$^1$
Osservatorio Astronomico di Trieste, Via G.B. Tiepolo 11,
I--34131, Trieste, Italy \\
$^2$ Dipartimento di Astronomia, Universit\`a
degli Studi di Trieste, I--34131, Trieste, Italy \\}

\date{Accepted: January 1995}
\maketitle

\begin{abstract}
Lithium abundances
in a sample of  halo dwarfs have been
redetermined by using the new T$_{eff}$ derived by Fuhrmann et al (1994)
 from  modelling of
the Balmer lines. These T$_{eff}$ are reddening independent, homogeneous  and
 of higher quality than those
based on broad band photometry.
Abundances have been derived  by generating new atmospheric models by using
the ATLAS-9 code by Kurucz (1993) with  enhanced $\alpha$-elements and
without the overshooting option. The revised abundances show
a remarkably flat {\it plateau} in the Li-T$_{eff}$
 plane for T$_{eff}$$>$ 5700 K with no evidence of trend with T$_{eff}$ or
falloff at the hottest edge. Li abundances are not
correlated with metallicity for [Fe/H]$<$ -1.4 in contrast with Thorburn
(1994).
All
the determinations are consistent with the same pristine lithium
abundance and the errors estimated for
individual stars fully account for the observed dispersion.
 The  weighted average Li value for the 24 stars of the plateau with
T$_{eff}$$>$ 5700 K and
[Fe/H]$\le$ -1.4,    is [Li] = 2.210 $\pm$ 0.013, or 2.224
  when non-LTE corrections by
Carlsson et al (1994) are considered.

\keywords{Stars: abundances -- Stars: Population II -- Stars: fundamental
parameters -- Galaxy: halo -- Cosmology: observations}
\end{abstract}

\section{Introduction}

The lithium observed in the atmospheres of unevolved halo stars is
generally believed to be an essentially unprocessed element which
reflects  the primordial yields.
In the framework of the standard BBN it  provides a sensitive measure of
$\eta$=$n_{b}/n_{\gamma}$ at the epoch of the primordial nucleosynthesis
and thus of the present
baryon density $\Omega_{b}$.
The primordial nature of the lithium of the halo dwarfs is inferred
from the presence of a constant lithium abundance for all the halo dwarfs
where convection is not effective (T$_{eff} \ge$ 5600 K).
Such an  uniformity is taken as evidence for the absence of any
stellar depletion during
the formation and the long life of the halo stars and
also as evidence for the absence of any production mechanism
acting either before or at the same time of
the formation of the halo population.

The existence of a real {\it plateau}
has been recently questioned
by Thorburn (1994), Norris et al (1994) and  Deliyannis et al (1993).
Thorburn (1994) found trends of the Li abundance both with
T$_{eff}$ and [Fe/H], while
Norris et al (1994) found  that the most extreme metal poor stars
provide  lower abundances by $\approx$ 0.15 dex, thus questioning
their genuine primordial value.
An intrinsic dispersion of Li abundances
in the plateau was claimed by Deliyannis et al (1993) from the analysis of the
{\it observable} EW and (b-y)$_{0}$ .
 These results open the possibility of
 substantial depletion
by rotational mixing where a certain degree of dispersion is foreseen for
different initial angular momenta of the stars and/or to a significant
Galactic lithium enrichment within the first few Gyrs.
Thus it appears rather problematic to pick up the precise primordial value from
the observations
of the Pop II stars.
Thorburn (1994) has suggested to estimate it
from the surface lithium abundances
of the hottest and  most metal-poor stars.

In this work we tackle these problems by recomputing
the lithium abundances for a significant subset  of
those stars already studied in literature for which
new and better effective temperatures are now available.
A possible
origin of the
systematic differences in the lithium abundances resulting in the most recent
determinations will also be discussed. Further details can be found in
Molaro et al (1995).

\section{Lithium Abundances}

\subsection{The role of T$_{eff}$ and of atmospheric models}

In the atmospheres of G dwarfs
lithium  is mainly ionized  and
 the abundance determination
is particularly sensitive to
the effective temperature and to
the T($\tau$) behaviour inside  the photosphere since it
requires large ionization corrections.
Conversely, lithium abundance is not particularly sensitive to
the stellar surface gravity and to the metallicity  of the star. Also
the LiI 6707 \AA~ line  is not generally saturated due to the intrinsically
low lithium abundance of the halo stars, and
therefore it is insensitive to the value of the microturbulent velocity.
The effective temperature is by far the most important parameter and
its accuracy determines the ultimate lithium abundance accuracy.
Unfortunately, the determination of the effective temperature for cool
 stars and in particular for metal poor stars is rather poor, as
discussed in detail
in Fuhrmann et al (1994). Fuhrmann et al pointed out the severe limitations
of the  methods based on broad band photometry, which they considered
inadequate
to provide accurate effective temperatures for individual stars.
The main arguments rely on the dependence of colour-based
temperatures on the
reddening corrections and on the particular color used.

\begin{table}
\caption{Table 1}
\begin{flushleft}
\begin{tabular}{lllrr} \hline
Star & T$_{B}\pm\sigma_{T_{B}}$ & $\log g$ & EW$\pm\sigma_{EW}$ &
[Li]$\pm\sigma_{Li}$ \\ \hline
           &              &      &            &                \\
HD   3567  & 5750$\pm$200 & 4.0 & 45$\pm$5.8 & 2.221$\pm$0.197 \\
HD  19445  & 6040$\pm$52  & 4.2 & 33.6$\pm$0.5 & 2.253$\pm$0.039 \\
HD  64090  & 5499$\pm$56  & 4.1 & 12.1$\pm$0.7 & 1.331$\pm$0.062 \\
HD  74000  & 6211$\pm$44  & 4.5 & 25$\pm$3.2 & 2.215$\pm$0.073 \\
HD 108177  & 6090$\pm$77  & 4.3 & 30$\pm$1.3 & 2.229$\pm$0.063 \\
HD 116064  & 5822$\pm$72  & 3.6 & 30$\pm$2.5 & 2.015$\pm$0.073 \\
HD 140283  & 5814$\pm$44  & 3.6 & 46.5$\pm$0.6 & 2.262$\pm$0.036 \\
HD 160617  & 5664$\pm$84  & 3.5 & 42$\pm$3.8 & 2.103$\pm$0.087 \\
HD 166913  & 5955$\pm$109 & 3.3 & 40$\pm$3.8 & 2.294$\pm$0.100 \\
HD 188510  & 5500$\pm$220 & 4.0 & 18$\pm$3.4 & 1.516$\pm$0.222 \\
HD 189558  & 5573$\pm$92  & 4.0 & 42$\pm$1.6 & 2.021$\pm$0.086 \\
HD 193901  & 5700$\pm$109 & 4.0 & 30$\pm$3.3 & 1.962$\pm$0.111 \\
HD 194598  & 5950$\pm$100 & 4.0 & 27$\pm$0.7 & 2.094$\pm$0.076 \\
HD 200654  & 5522$\pm$119 & 3.2 &  8$\pm$1.7 & 1.129$\pm$0.143 \\
HD 201889  & 5645$\pm$61  & 4.1 &  5$\pm$3.3 & 1.065$^{+0.230}_{-0.500}$ \\
HD 201891  & 5797$\pm$57  & 4.4 & 24.3$\pm$0.8 & 1.925$\pm$0.048 \\
HD 211998  & 5338$\pm$65  & 3.5 & 13$\pm$3.4 & 1.219$\pm$0.150 \\
HD 219617  & 5815$\pm$76  & 4.2 & 40.2$\pm$0.8 & 2.198$\pm$0.062 \\
BD  2$^{\circ}$ 3375 & 6034$\pm$60  & 4.0 & 31.5$\pm$2.1 & 2.198$\pm$0.058 \\
BD  3$^{\circ}$  740 & 6264$\pm$73  & 3.5 & 17.3$\pm$1.3 & 2.062$\pm$0.065 \\
BD  9$^{\circ}$  352 & 6285$\pm$77  & 4.5 & 34$\pm$6.7 & 2.429$\pm$0.130 \\
BD  9$^{\circ}$ 2190 & 6452$\pm$60  & 4.0 & 18$\pm$3.4 & 2.200$\pm$0.107 \\
BD 17$^{\circ}$ 4708 & 6100$\pm$110 & 4.1 & 25$\pm$1.7 & 2.139$\pm$0.084\\
BD 21$^{\circ}$  607 & 6135$\pm$70  & 4.0 & 25$\pm$2.6 & 2.190$\pm$0.074 \\
BD 23$^{\circ}$ 3130 & 5190$\pm$84  & 2.7 & 13$\pm$1.3 & 1.066$\pm$0.095\\
BD 24$^{\circ}$ 1676 & 6278$\pm$76  & 3.9 & 27$\pm$1.6 & 2.296$\pm$0.062 \\
BD 26$^{\circ}$ 2606 & 6161$\pm$64  & 4.1 & 30$\pm$1.2 & 2.252$\pm$0.050 \\
BD 29$^{\circ}$  366 & 5760$\pm$64  & 3.8 & 14$\pm$2.9 & 1.641$\pm$0.105 \\
BD 37$^{\circ}$ 1458 & 5451$\pm$59  & 3.5 & 11$\pm$2.1 & 1.226$\pm$0.107 \\
BD 38$^{\circ}$ 4955 & 5337$\pm$73  & 4.5 &  8$\pm$3.3 & 1.016$\pm$0.250 \\
BD 42$^{\circ}$ 3607 & 5836$\pm$66  & 4.4 & 47$\pm$4.5 & 2.298$\pm$0.083 \\
BD 66$^{\circ}$  268 & 5511$\pm$91  & 4.0 & 10$\pm$7.8 & 1.237$\pm$0.476 \\
G 64-12    & 6356$\pm$75  & 3.9 & 25.8$\pm$2.4 & 2.318$\pm$0.074 \\
G 64-37    & 6364$\pm$75  & 4.1 & 14$\pm$1.2 & 2.029$\pm$0.066 \\
G 66-9     & 5885$\pm$83  & 4.6 & 29$\pm$2.9 & 2.047$\pm$0.081 \\
G 206-34   & 6258$\pm$50  & 4.3 & 27$\pm$2.5 & 2.265$\pm$0.060 \\
G 239-12   & 6260$\pm$70  & 4.1 & 24$\pm$3.1 & 2.214$\pm$0.083 \\
G 255-32   & 5962$\pm$53  & 4.0 & 30$\pm$2.9 & 2.119$\pm$0.064 \\
LP 608 62  & 6435$\pm$52  & 4.1 & 21.5$\pm$2.3 & 2.282$\pm$0.063 \\
           &              &     &              &           \\ \hline
\end{tabular}
\end{flushleft}
\end{table}

Fuhrmann et al (1994) derived
T$_{eff}$ from the full spectral synthesis of the Balmer lines
for a large sample of stars. These T$_{eff}$ are
reddening independent and  they show a high degree of internal consistency when
the various
 members of the serie
are used. Being obtained from absorption lines, they  are particularly suitable
for line abundance applications. Fuhrmann  et al
are also able to provide errors in the T$_{eff}$ for individual stars and
in most cases they are as good as $\pm$ 50 K, which is
a factor 2 smaller
than the grossly estimated errors for photometric-based T$_{eff}$.

Out of the Fuhrmann et al' sample,  39 have already been studied for
lithium. They represent a significant fraction
 of the presently available lithium determinations,
and form an unique  sample with a good and homogeneous T$_{eff}$.

\begin{figure}
\psfig{figure=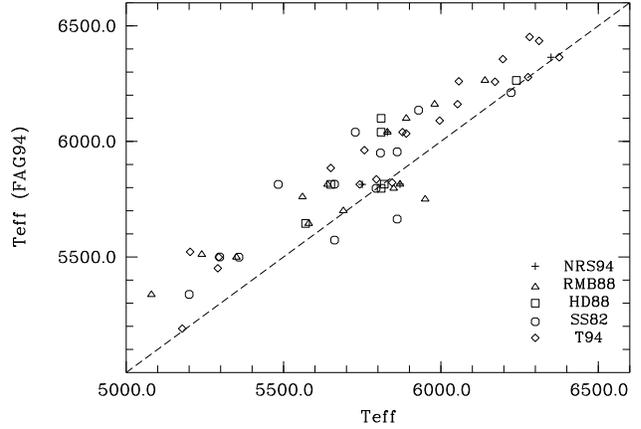,width=8.8cm,angle=-90}
\caption{T$_{eff}$ from Fuhrmann et al (1994) versus T$_{eff}$ from
Li literature. }
\end{figure}

\begin{figure}
\psfig{figure=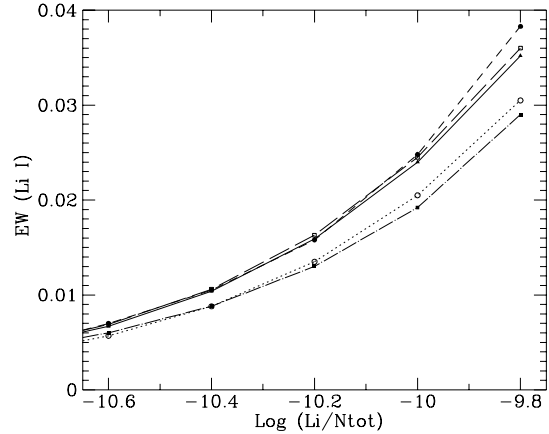,width=8.8cm,clip=t}
\caption{ Curves of Growth for different models. Dash: Bell;
 solid line: ATLAS8;
dot: ATLAS9; Long dash: ATLAS9 without overshooting;
dot - long dash: Kurucz 1991 adopted
by Thorburn 1994. }
\end{figure}

On average, the Fuhrmann et al temperatures are
125$\pm$120 K higher than those previously used in
literature on lithium (cfr Fig. 1).
The presence of considerable scatter together with a systematic offset
suggest that there are intrinsic differences in the individual temperatures
derived by various methods. The Li equivalent widths
have been taken from the literature  and the
theoretically derived random errors from
Deliyannis et al (1993) or computed following their
prescriptions. In the case of multiple measurements of the lithium line
of the same star we adopted the weighted average to minimize the errors.
Atmospheric models may be important for the lithium abundance.
 The role played by different atmospheric models is illustrated in Fig. 2
where several curves of growth obtained with different atmospheric models
are shown.
 The COGs are for T$_{eff}$ = 6000 K, $\log$g = 3.5,
microturbulence $\xi$ = 2 km s$^{-1}$ and [Fe/H]=-3.0,
but a similar behaviour
 is shown by other temperatures
relevant for Li.
The curves
of growth for lithium
obtained by using ATLAS-9 and ATLAS-8 are notably different.
In Fig. 2 are also shown the COG used by  Thorburn (1994)
referred   to unpublished models of Kurucz (1991)
and to those of Bell and Gustaffsson.
For a given EW
the abundances of the most recent Kurucz codes
are $\approx$ 0.1 dex higher than all the other curves.
We have computed a grid of atmospheric models
where the convection is treated with the mixing length
theory but without the overshooting option. The corresponding COG,
also shown in Fig. 2, is very close to the COG of
the old Kurucz  or Bell and Gustaffsson models. This shows that
 the implementation of the approximate treatment
for overshooting in the
ATLAS-9 code is likely to be responsible for the difference among the
 versions of the ATLAS code.
 The center of the convective bubbles stops at the top of
the convective zone so that convective flux extends one bubble radius
above the end of the convection zone.
Overshooting  rises the temperature in correspondence of the depths
relevant for the formation of the lithium line, thus  resulting
in higher abundance for the same equivalent width.
The effect is
almost negligible at solar metallicities but it
increases towards lower metallicities where
the fraction of the total flux transported via convection
increases.

The presence of abundances derived with different atmospheric models is
also responsible for the
systematic differences in the lithium abundances
for common stars among different authors. This factor has been essentially
overlooked
and has introduced  a spurious scatter in a
straightforward compilation of the literature data. The
comparison of the measurements of Thorburn (1994) with those of the literature
shows that the equivalent widths are almost the same but the Thorburn
abundances derived using Kurucz (1991) are systematically higher than those
by other authors.
The finding by Norris et al (1994) showing that Li abundances in stars
with [Fe/H]$<$-3.0 are on average about 0.15 dex lower than the higher
metallicity halo stars from Thorburn (1994) can be also understood
as a model effect.
Norris computed the Li abundances by using  Bell and Gustaffsson
 atmospheric
models which give  abundances lower than those by   Thorburn (1994).
The comparison between the stars studied by both Norris et al (1994)
and Thorburn (1994)
shows nearly identical
EWs for both, but lower abundances in the former authors.

Implementation of overshooting improves the reproduction of the
solar spectrum, but the effects on metal poor stars
have not been yet fully tested and it should be considered with caution until
a deeper analysis of possible secondary effects on metal poor objects
is carried out. Here we have not used this option and the
new lithium abundances have been determined
by computing Kurucz (1993) atmospheric models
without overshooting.
For all the stars  the models have been computed
with the temperatures available from Fuhrmann et al (1994).
The gravities used are taken from
the literature according to the compilation
of Deliyannis et al (1993) or computed from {\it} c$_{0}$ colors, and
are reported in Table 1.
Convection is treated with
the mixing length theory with a scale height over pressure scale of
1.25. The choice of this parameter is not critical
and a change from 0.5 to 2 produces almost negligible effects on the
lithium abundance.
If we use the Kurucz 1993 grid which includes overshooting
the lithium abundance is
increased by $\approx$ 0.09 dex, but all other conclusions are still
valid, since they are not dependent on this assumption.
In view of the important implications related to the primordial abundances,
it is desirable that a deeper discussion may follow shortly.

\begin{figure}
\psfig{figure=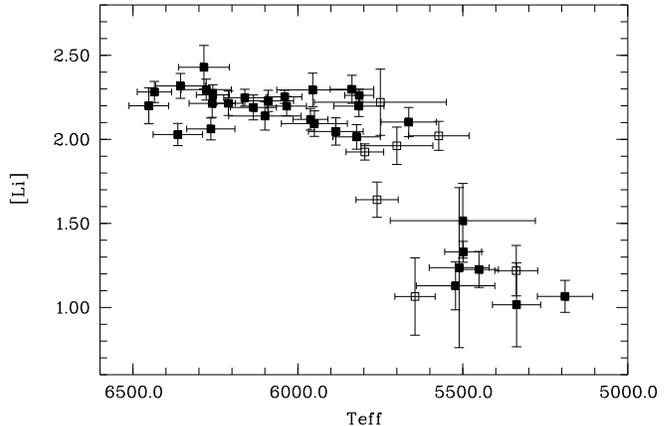,width=8.8cm,angle=-90}
\caption{[Li] versus T$_{eff}$. Open squares are for stars with metallicities
in the range -1.4$<$[Fe/H]$<$-1.0.
}
\end{figure}

\section{Results}

The Li abundances  in the Li-T$_{eff}$ plane with associate 1 $\sigma$ errors
are shown in Fig. 3.
 Following Rebolo et al (1988)
and Deliyannis et al (1990), stars with -1.4$<$[Fe/H]$<$-1.0 (open squares in
Fig. 3)
are taken off from  sample
 of genuine halo stars  since they may show some depletion.
The errors shown in Fig. 3 are those given in the last column of Table 1.
They have been estimated by summing under
quadrature the errors in the lithium abundance  produced by the uncertainties
in T$_{eff}$ and in EW, also given in Table 1, namely
$\sigma_{Li}^2 = \sigma_{Li}^2(EW) + \sigma_{Li}^2(T_{eff})$. We have
 considered negligible
the effects produced by uncertainties in gravity, microturbulence and
metallicity.
The lithium abundances show a plateau extending up to nearly 6500 K, with no
evidence of falloff at the hottest edge as expected by microscopic diffusion
models,
and with the depletion
region bending at T$_{eff}$ $\approx$ 5700 in good agreement with what observed
in the Hyades.

The existence of a tilt of the plateau has received
considerable attention (Molaro 1987, Rebolo et al 1988,
Thorburn 1994, Norris et al 1994) since the
 presence of a moderate stellar
depletion in the cool edge {\it plateau} implies that only
the highest values are close to the pristine value.
Norris et al (1994) and Thorburn (1994)
found a slope of 0.03 and 0.024 for 100 K, respectively.
Thorburn (1994) claimed also that when this underlying trend with T$_{eff}$ is
taken into account the increase of N(Li) with the metallicity becomes
notable already at
[Fe/H]$\approx$-2.0.
Considering only stars with Teff $>$ 5700 K and [Fe/H]$\le$ -1.4 the
 weighted  linear  fit
is [Li]$\propto 0.58(\pm 0.88) \cdot (T_{eff}/10^4)$. The  slope we
 found is one order of magnitude lower than those of
Norris et al (1994) or Thorburn (1994), and is consistent with
a real {\it plateau} in the undepleted region of the Li-T$_{eff}$ plane.
Non-LTE effects have been studied by Carlsson et al (1994), who provide
corrections
of 0.020, 0.015  and 0.010 at 5500, 6000 and 6500 K respectively,
and for [Fe/H] between -1.0 and -3.0. Once we correct our abundances for
nonLTE effects  the
slope of the fit becomes even smaller: 0.29($\pm$ 0.93).

The lithium abundances versus the stellar metallicity are shown in Fig. 4
and do not show any clear trend with iron over two orders of magnitude of
increase in the stellar metallicity. Strictly speaking we observe a
 decrease in the  lithium abundance with the increasing of the metallicity.
The weighted regression analysis
gives a negative slope with lithium decreasing by 0.008 dex for
$\Delta$[Fe/H]=1. By contrast Thorburn (1994) found an increase
in Li by 0.4 dex from the [Li]=2.20 at [Fe/H]=-3.5 up to 2.60 at
[Fe/H]=-1.0.
Our  interception  at [Fe/H]=-3.5 is of [Li]=2.215, fully consistent with the
value of [Li]=2.221 obtained from the hottest halo dwarfs
(T$_{eff}$=6400 K). In our data sample both the lowest
metallicity stars, i.e. the oldest, and the hottest subdwarf, i.e.
the less depleted, share the same Li abundance.

Fig. 5 shows a zoom of the plateau region with the errors in the Li
abundances at the 3 $\sigma$ level plotted for each star.
On the plateau all the Li abundances are
consistent  with a unique pristine Li abundance. In 10 out of 24
cases the consistency is achieved already at 1  $\sigma$ confidence level.
For three stars (G 64-13, BD 3$^{\circ}$ 740, HD 116064)
the full  3$\sigma$ error box is required
to achieve the consistency, and they  might show a real dispersion if errors
can be further reduced.
With the present estimated errors
our analysis does not support the presence of real dispersion
on the plateau region, and it
seems likely that the dispersion claimed by
Thorburn (1994), or the correlations of Li with metallicity or temperature,
are artifacts caused by errors in the effective temperatures.
We stress that both the absence of a tilt and  dispersion in the
plateau region are independent of the assumption we made on overshooting, since
its inclusion would
increase the Li abundance by about the same amount for all the
stars.
The weighted mean on the plateau of the 24 stars with T$_{eff}$
$\ge$ 5700 and [Fe/H]$\le$ -1.4, where each
abundance is weighted inversely by its own variance in the sum,
 is [Li]=2.210 $\pm$ 0.013.
When the non-LTE corrections of Carlsson et al (1994) are considered,
the mean rises to 2.224 $\pm$ 0.013.
This value is somewhat higher than the 2.08$\pm$0.1
previously  estimated by Spite and Spite (1982),
 Hobbs and Duncan (1988), Rebolo et al (1988), and Molaro (1991).
The increase in the value of the plateau results
from the increase of the Fuhrmann et al effective
temperatures compared to those previously used in the lithium
literature.

The present analysis shows that when very precise effective temperatures
and individual errors are considered, the Li  abundances on the
plateau show no trends either with
T$_{eff}$ in a range of 600 K or with the stellar metallicity over two
orders of magnitude. The lithium abundances are all closely gathered
and are consistent with the same
initial abundance, thus confirming
that the lithium observed in these stars is
essentially undepleted and very close to the primordial value
as already put forward by
Spite and Spite (1982).

\begin{figure}
\psfig{figure=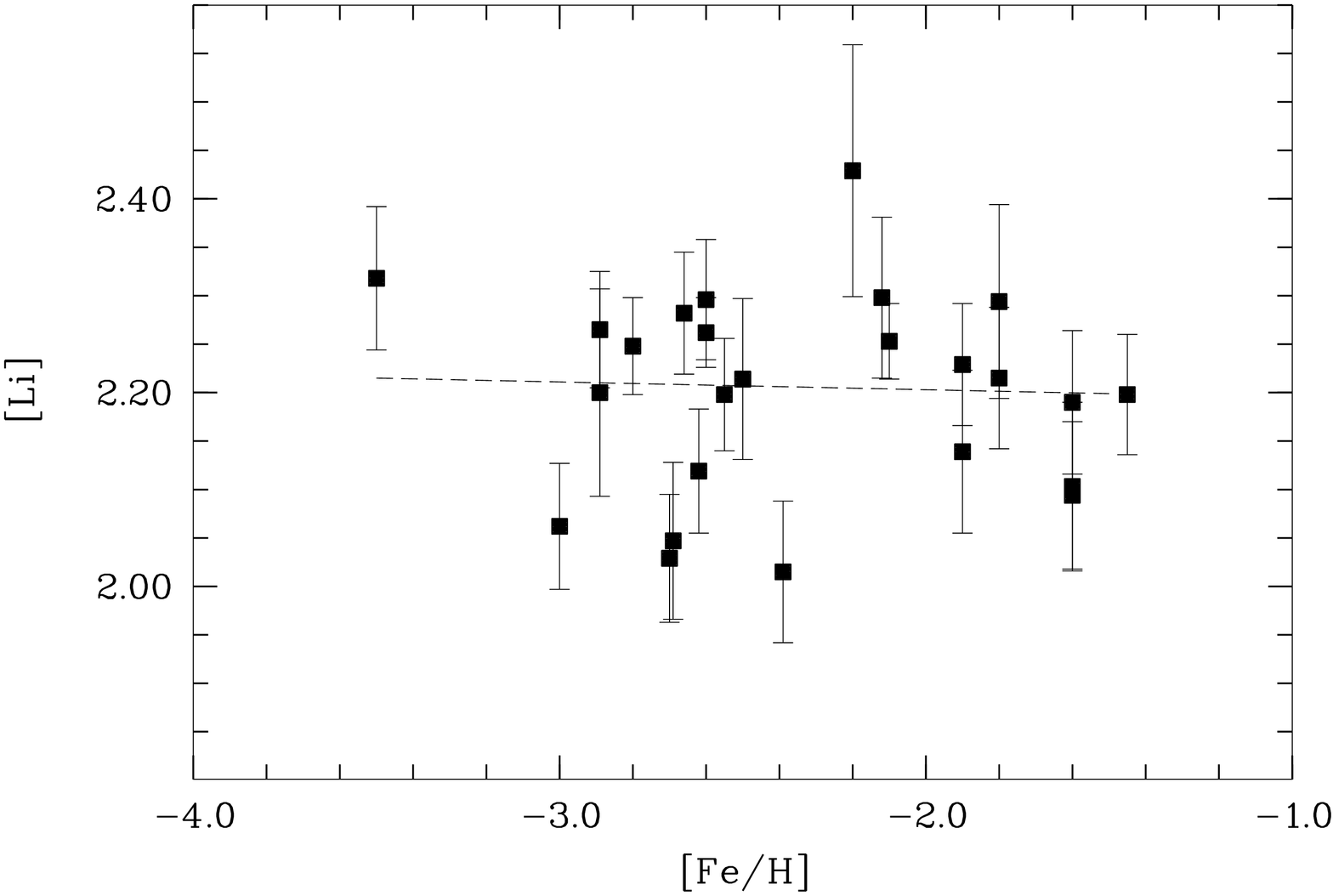,width=8.8cm,angle=-90}
\caption{Li-[Fe/H].}
\end{figure}

\begin{figure}
\psfig{figure=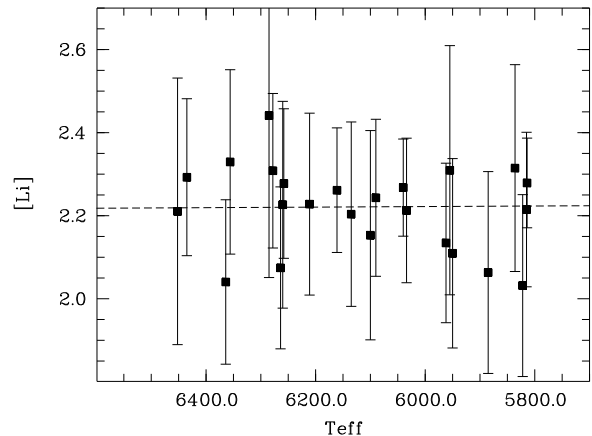,width=8.8cm,angle=-90}
\caption{Zoom on the {\it plateau}, with the [Li] abundances corrected for
non-LTE effects.  The errors are at 3 $\sigma$. The dash line shows the
weighted mean at [Li]=2.224. }
\end{figure}

\acknowledgements{It is a pleasure to thank Fiorella Castelli and
Bob Kurucz for many useful discussions in particular on the role played by
atmospheric models on the lithium abundance. Partial support from
a bilateral CNR grant is acknowledged.}

\end{document}